\documentclass[
reprint,
superscriptaddress,
 amsmath,amssymb,
 aps,
]{revtex4-2}

\usepackage{graphicx}
\usepackage{dcolumn}
\usepackage{bm}
\usepackage{xcolor} 
\usepackage{lipsum} 
\usepackage{verbatim}   
\usepackage{hyperref}
\usepackage{gensymb}
\usepackage{textgreek}

\begin{document}

\preprint{APS/123-QED}

\title{Generation of continuous-wave laser light at 148.4 nm using cavity-enhanced second harmonic generation in BaMgF\textsubscript{4}}

\author{Keerthan~Subramanian}
\email{keerthan.subramanian@uni-mainz.de}
\affiliation{QUANTUM, Institut für Physik, Johannes Gutenberg-Universität Mainz, 55128 Mainz, Germany\\}

\author{Hiroki~Tanaka}
\affiliation{Leibniz-Institut für Kristallzüchtung (IKZ), 12489 Berlin, Germany\\}

\author{Simon~J.~Herr}
\affiliation{Fraunhofer-Institut für Physikalische Messtechnik IPM, 79110 Freiburg, Germany\\}

\author{Nutan~Kumari~Sah}
\affiliation{QUANTUM, Institut für Physik, Johannes Gutenberg-Universität Mainz, 55128 Mainz, Germany\\}
\affiliation{Deutsches Elektronen-Synchrotron (DESY), 15738 Zeuthen, Germany}
\affiliation{Department of Physical Sciences, Indian Institute of Science Education and Research Mohali 140306, India}

\author{Gaurav~Jha}
\affiliation{QUANTUM, Institut für Physik, Johannes Gutenberg-Universität Mainz, 55128 Mainz, Germany\\}
\affiliation{Department of Physics, Indian Institute of Technology Kanpur, Kanpur 208016, India\\}

\author{Florian~Zacherl}
\author{Srinivasa~Arasada~Pradeep}
\affiliation{QUANTUM, Institut für Physik, Johannes Gutenberg-Universität Mainz, 55128 Mainz, Germany\\}

\author{Valerii~Andriushkov}
\affiliation{Helmholtz Institut Mainz, 55128 Mainz, Germany}
\affiliation{GSI Helmholtzzentrum für Schwerionenforschung GmbH, 64291 Darmstadt, Germany}

\author{Ke~Zhang}
\affiliation{QUANTUM, Institut für Physik, Johannes Gutenberg-Universität Mainz, 55128 Mainz, Germany\\}
\affiliation{Center for Photonic Quantum Precision Measurement, Advanced Research Institute of Multidisciplinary Science, Beijing Institute of Technology, Beijing 100081, China}
\affiliation{Beijing Institute of Technology, State Key Laboratory of Environment Characteristics and Effects for Near-space, Beijing 100081, China}

\author{Darius~Fenner}
\affiliation{QUANTUM, Institut für Physik, Johannes Gutenberg-Universität Mainz, 55128 Mainz, Germany\\}

\author{Yumiao~Wang}
\affiliation{QUANTUM, Institut für Physik, Johannes Gutenberg-Universität Mainz, 55128 Mainz, Germany\\}
\affiliation{Key Laboratory of Nuclear Physics and Ion-beam Application (MoE), Institute of Modern Physics, Fudan University, 200433 Shanghai, China\\}

\author{Milena~Hugenschmidt}
\affiliation{Fraunhofer-Institut für Physikalische Messtechnik IPM, 79110 Freiburg, Germany\\}

\author{Frank Kühnemann}
\affiliation{Fraunhofer-Institut für Physikalische Messtechnik IPM, 79110 Freiburg, Germany\\}
\affiliation{Physikalisches Institut, Universität Freiburg, 79104 Freiburg, Germany\\}

\author{Gaetano~G.~M.~Bonetti}
\author{Shoichi~Ui}
\author{Matthias~Bickermann}
\affiliation{Leibniz-Institut für Kristallzüchtung (IKZ), 12489 Berlin, Germany\\}

\author{Chenxi~Ma}
\author{Xian~Zheng}
\affiliation{Institut für Festkörperphysik, Leibniz Universität Hannover, 30167 Hannover, Germany}
\author{Michael~Zopf}
\affiliation{Institut für Festkörperphysik, Leibniz Universität Hannover, 30167 Hannover, Germany}
\affiliation{Laboratorium für Nano- und Quantenengineering, Leibniz Universität Hannover, 30167 Hannover, Germany}

\author{Bettina~Lommel}
\affiliation{GSI Helmholtzzentrum für Schwerionenforschung GmbH, 64291 Darmstadt, Germany}

\author{Jan~C.~Müller}
\affiliation{QUANTUM, Institut für Physik, Johannes Gutenberg-Universität Mainz, 55128 Mainz, Germany\\} 

\author{Stephan~Hannig}
\affiliation{Agile Optic GmbH, 38110 Braunschweig, Germany}

\author{Dmitry~Budker}
\affiliation{QUANTUM, Institut für Physik, Johannes Gutenberg-Universität Mainz, 55128 Mainz, Germany\\}
\affiliation{Helmholtz Institut Mainz, 55128 Mainz, Germany}
\affiliation{GSI Helmholtzzentrum für Schwerionenforschung GmbH, 64291 Darmstadt, Germany}
\affiliation{University of California, Berkeley, California 94720, USA\\}

\author{Ferdinand~Schmidt-Kaler}
\affiliation{QUANTUM, Institut für Physik, Johannes Gutenberg-Universität Mainz, 55128 Mainz, Germany\\}
\affiliation{Helmholtz Institut Mainz, 55128 Mainz, Germany}

\author{Lars~von~der~Wense}
\affiliation{QUANTUM, Institut für Physik, Johannes Gutenberg-Universität Mainz, 55128 Mainz, Germany\\}

\date{\today}

\begin{abstract}
The development of a nuclear optical clock based on \textsuperscript{229}Th is expected to influence the next generation of time and frequency standards. At its heart is the need for a continuous wave (CW) at laser at 148.4~nm to drive the clock transition between the ground and excited isomeric state. In this work we experimentally investigate the potential of BaMgF\textsubscript{4} crystals to create a  CW solid state laser at the vacuum ultraviolet (VUV) wavelength of 148.4~nm via cavity-enhanced second harmonic generation. For this purpose, a BaMgF\textsubscript{4} crystal was grown, optically polished and periodically poled. The crystal was inserted into a power-enhancement cavity, resonant at the fundamental wavelength of 296.8~nm and the generated laser light at 148.4~nm was characterized. Within this proof-of-concept investigation, a VUV output power of typically ($16\pm1$)~pW is obtained. This marks the first time that this type of crystal is used to generate VUV laser light. The experimental findings are compared to theoretical expectations and provide a clear path for future improvements, which could have implications not only for compact laser sources in the VUV but also solid-state as well as single-ion based \textsuperscript{229}Th nuclear clocks. 
\end{abstract}

\maketitle
\nocite{*}
\section{Introduction}
The history of timekeeping has been marked by a search for references with high quality factors whose oscillation frequency is less susceptible to external perturbations. With the introduction of the Cs microwave standard for the definition of a second, a highly reproducible time and frequency standard with high fractional frequency stability was achieved. Present-day optical atomic clocks have long surpassed the stability of the Cs standard making the redefinition of the SI second imminent \cite{dimarcq2024roadmap}. A large fraction of the remaining systematic frequency uncertainties as well as instabilities of these clocks is due to the interaction of the atomic reference with the external environment \cite{ludlow2015optical}. While there have been complementary approaches to developing the next generation of atomic clocks \cite{kozlov2018highly}, the possibility to build a clock based on a nuclear transition \cite{peik2003nuclear} has attracted increasing attention.

At the center of such an endeavor is the \textsuperscript{229m}Th nuclear isomer with an anomalously low energy of about 8.4~eV above the ground state bestowing it a unique position in the entire nuclear energy landscape \cite{von2016direct}. The small electric and magnetic moments of the nucleus resulting in orders of magnitude better insensitivity to external perturbations as well as the prospect of developing an all solid-state clock \cite{peik2003nuclear,rellergert2010constraining} have made the \textsuperscript{229m}Th nuclear transition a formidable candidate for the next generation of time and frequency standards. A comprehensive review on the history of \textsuperscript{229m}Th can be found in Ref.~\cite{vonderwense2020review}.

The first laser excitation of \textsuperscript{229m}Th was achieved in 2024 based on the irradiation of \textsuperscript{229}Th-doped crystals \cite{tiedau2024laser,elwell2024laser}. These experiments used pulsed lasers based on four-wave mixing in xenon, which have the advantage of high pulse energies of 1 to 15~\textmu J per pulse at 30~Hz repetition rate, however, at the cost of a broad spectral bandwidth of about 10~GHz, making them unsuitable for the development of a nuclear clock. These experiments were soon followed by direct frequency comb spectroscopy, using a VUV frequency comb generated via intra-cavity high-harmonic generation \cite{zhang2024frequency}. In this experiment, a single comb-mode was used for the spectroscopy, resulting in a narrow spectral linewidth. However, the remaining $\sim10^{5}$ comb-modes remain unused, making the process inefficient.

\section{CW laser \MakeLowercase{at} 148.4~\MakeLowercase{nm}}
CW lasers at 148.4 nm make the ideal light sources for driving the \textsuperscript{229}Th transition in a nuclear clock. In this study, generation of CW VUV laser light at 148.4~nm using cavity-enhanced second-harmonic generation (SHG) in periodically poled BaMgF\textsubscript{4} (PPBMF) is demonstrated for the first time.
 Recently, two other groups have succeeded in the development of CW 148.4 nm laser sources: A solid-state laser source was demonstrated by frequency doubling in a random quasi-phase-matched SrB\textsubscript{4}O\textsubscript{7} (SBO) crystal, achieving a power of about 1~nW \cite{lal2025continuous}, which has been used for continuous-wave laser spectroscopy of \textsuperscript{229m}Th \cite{morawetz2026cwspectroscopy} and the development of a nuclear optical clock \cite{Toscani2026}. An even larger power of about 10 \textmu W was obtained via four-wave mixing in cadmium vapor \cite{xiao2026continuous} and has as well been used for nuclear optical clock development \cite{Huang2026}.

Developing a high power, all-solid-state CW laser system at 148.4~nm still remains an open quest. The most straight-forward approach is to perform SHG in a suitable nonlinear crystal. Due to the requirements of $\chi^{(2)}$-nonlinearity as well as VUV transparency, only a few crystals appear suitable for this approach. Following a compilation found in \cite{shao2020pushing}, the most promising candidates are BPO\textsubscript{4}, SrB\textsubscript{4}O\textsubscript{7}, SrMgF\textsubscript{4} and BaMgF\textsubscript{4}. However, since none of these crystals is amenable to birefringent phase-matching for frequency-doubling from 296.8~nm to 148.4~nm, quasi-phase-matching (QPM) is required. The commonly applied method for QPM is by introducing periodic domain inversions into the crystal by applying an external electric field, known as electric-field periodic poling, which requires the nonlinear crystal to be ferroelectric. Of the crystals mentioned above, only BMF has been confirmed to be ferroelectric and is therefore the only crystal known to be suitable for QPM to 148.4 nm via the established technique of electric-field periodic poling.
 BMF has long been discussed for such applications \cite{buchter2001periodically} and was already used for frequency doubling, with 368 nm being the shortest wavelength previously obtained \cite{villora2009birefringent}. However, the largest nonlinear coefficient of BMF was estimated to be 0.039~pm/V (tensor element $d$\textsubscript{32}), which is about two orders of magnitude smaller than for crystals typically used in nonlinear frequency doubling applications into the UV spectral range. With more than 1.5~pm/V, non-ferroelectric SBO offers a larger nonlinear coefficient, which was already made use of in the process of random quasi-phase-matching, where random domain inversion was introduced into an SBO crystal during the growth process \cite{zaitsev2008domain} and a particularly suitable spot on the crystal was used \cite{lal2025continuous}. The non-deterministic nature of this process makes it hard to replicate such crystals on a larger scale. To take full advantage of the larger nonlinear coefficients of the non-ferroelectric crystals, researchers are exploring novel methods to achieve deterministic quasi-phase-matching\cite{shao2020pushing,perlov2024method,vasilyev2026148nm}.

The potential of PPBMF to generate CW laser light at 148.4~nm is experimentally investigated. Compared to the Cd-Vapor source, our approach with future power scaling could be a more compact solution with the possibility of integrating both the laser source as well as the Th nuclei in a single crystal\cite{morgan2025proposal}. In addition, even though the powers we achieve in this proof-of-principle demonstration are smaller than the all-solid state SBO approach, in terms of the potential for deterministically scaling up power our approach provides a viable solution. Despite SBO having a nearly 30x higher $\chi^{(2)}$, the current demonstration\cite{lal2025continuous} makes use of random QPM exploiting the spatial fourier component of $\chi^{(2)}$ at the coherence length $L_c$. This is completely random and not reproducible across the transverse extent of a single sample let alone across multiple samples. Combined with its non-ferroelectric properties making electric field poling impossible, a clear path to deterministic power scaling with SBO currently does not exist . For our proof-of-principle demonstration, VUV transparent and homogeneous BMF crystals were grown, optically polished and periodically poled. A PPBMF crystal was inserted into an enhancement cavity, operational at the fundamental wavelength of 296.8~nm. This was needed in contrast to the single-pass experiment with SBO since the $\chi^{(2)}$ is orders of magnitude smaller and higher powers are needed to exploit the quadratic scaling of the SHG output.
The generated VUV light was coupled out from the cavity and detected using either a micro-channel plate (MCP) or a photo-multiplier tube (PMT) for measuring the SHG power. In the following, each step will be described individually.

\section{Crystal Growth}
\begin{figure}
    \centering
    \includegraphics[width=0.8\linewidth]{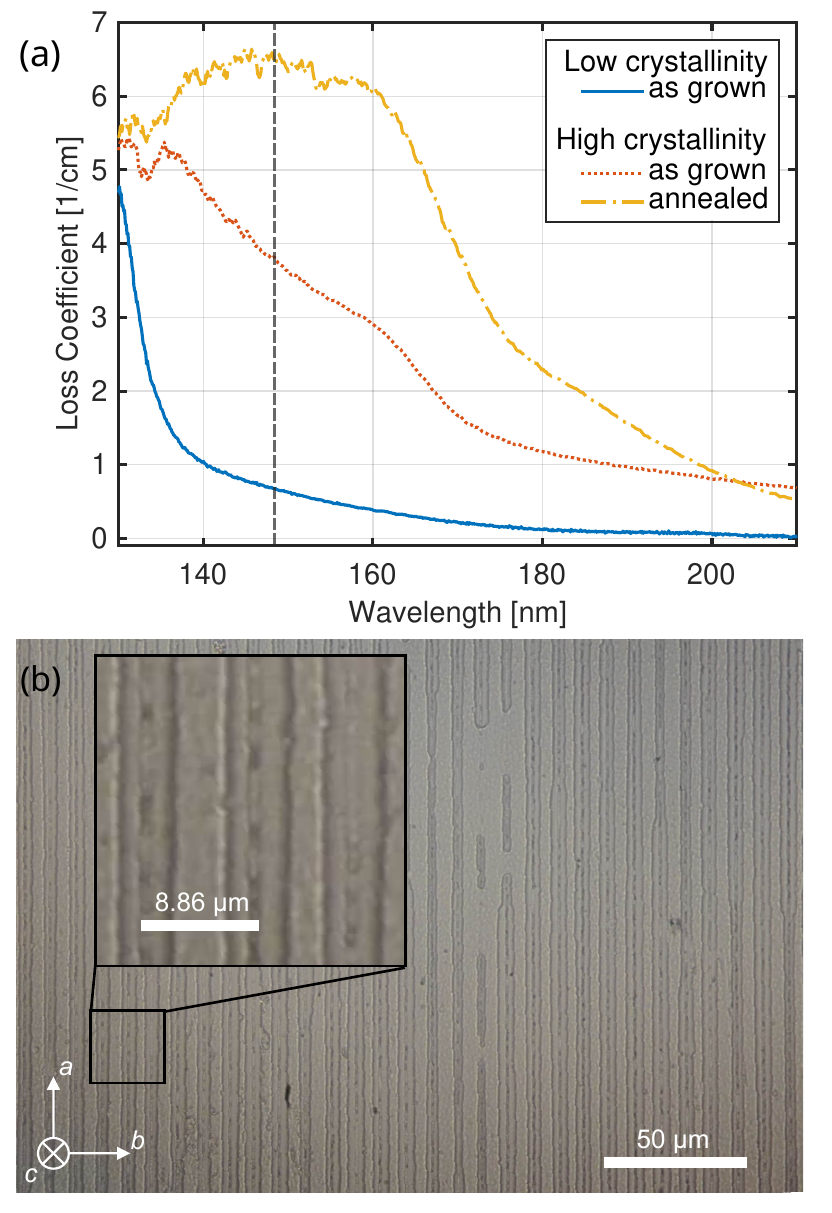}
    \caption{\textbf{BaMgF\textsubscript{4} for second harmonic generation.} BaMgF\textsubscript{4} (BMF) used for SHG is transparent for VUV wavelengths and is ferroelectric, making it amenable to quasi-phase matching (QPM) by periodic poling. (a) Loss coefficients obtained from the measured transmission spectra of different c-cut BMF samples. The lowest loss in the VUV spectral range is obtained for a crystal grown in aggressive fluorinating atmosphere (blue, solid) resulting in low crystallinity and lower poling performance. A sample of high crystallinity (red, dotted) results in higher loss coefficients (lower VUV transmission), but improved poling quality. Post-growth vacuum annealing (yellow, dash-dot) reduced the loss above 200 nm, but further increased VUV loss. The high-crystallinity annealed sample (yellow, dash-dot) was used for the SHG experiments. The dashed vertical line indicates the wavelength of the \textsuperscript{229m}Th nuclear transition.
    (b) Typical poling quality of the 300-\textmu m-thick, periodically poled BMF crystal used for this study with poling period of 8.86~\textmu m ({9\textsuperscript{th}} order QPM for d$_{31}$). The scale bar in the inset shows the distance between the center positions of two adjacent inverted domains. The domains are visualized using standard bright field microscopy after domain selective etching\cite{herr2023fanout}}
    \label{fig:01_BMF}
\end{figure}

BMF crystal growth was carried out at the Leibniz-Institut für Kristallzüchtung (IKZ, Berlin) using the Czochralski technique \cite{herr2023fanout}. Each crystalline boule was pulled from a stoichiometric melt in a graphite crucible along the crystallographic $c$-axis using an oriented seed crystal. As the starting materials, UV-grade BaF\textsubscript{2} and MgF\textsubscript{2} crystals (Korth Kristalle) were used.
The grown boules (approximately 20 mm in diameter and 50 mm in length) were then oriented, cut, and mechanically polished (with 1\textmu m polycrystalline diamond abrasives) to prepare samples with dimensions of (a $\times$ b $\times$ c) $5\times5\times0.3$ mm\textsuperscript{3} for periodic poling. 

Transmission spectra of $c$-cut (10-mm-long, 15 mm\textsuperscript{2} aperture) BMF samples were measured under vacuum using a VUV spectrometer (Optes) without a polarizer, and the resulting loss-coefficient spectra are shown in Fig.~\ref{fig:01_BMF}a. The Fresnel reflection losses are corrected using the reported Sellmeier equations \cite{villora2009birefringent}. In this work, we use crystallinity as a qualitative indicator of the crystal quality, evaluated from the width of the X-ray rocking curve. In general, a narrower rocking-curve peak indicates a smaller angular spread of the crystallographic planes and, thus, a higher degree of structural perfection. BMF samples grown in aggressive fluorinating atmosphere (blue, solid curve) showed a high VUV transparency (e.g., up to 67\% transmission at 148.4 nm for a 5-mm-thick crystal). However, due to their lower crystallinity compared to samples grown in non-fluorinating atmosphere, these crystals appear to be harder to periodically pole with small domain width. In contrast, crystals with higher crystallinity (red, dotted curve) exhibit a lower VUV transparency but showed superior periodic poling performance. As their low VUV transparency is partly attributed to scattering centers, we applied post-growth annealing in vacuum as reported in Ref.~\cite{zhao2011formation}. For annealing, the crystal was ramped to $910\degree$C within 12 hours, stayed at that temperature for 24 hours, before being ramped down within 24 hours. Although annealing was found to reduce scattering centers, this simultaneously increased the absorption loss for wavelengths below 200 nm (yellow, dash-dot curve). Although the mechanism of the UV absorption remains unclear, it may be associated with the formation of trace foreign crystallographic phases, which were not detected by X-ray diffraction, and/or color centers related to fluorine vacancies generated during vacuum annealing. Due to the superior poling quality, crucial for QPM, and reduced loss at the fundamental wavelength, the annealed high crystallinity BMF crystal (yellow, dash-dot curve) was used throughout this investigation. However, this was at the cost of VUV transmission, which was reduced to about $(3.6\pm1)$\%.

BMF is an orthorhombic, optically biaxial crystal. For different polarizations of the fundamental wave, we calculated the coherence lengths ($L_c = 2\pi/\Delta k$, where $\Delta k = k_2 - 2k_1$, $k_1$ and $k_2$ being the wavevectors of the fundamental and second-harmonic) using the reported Sellmeier coefficients\cite{villora2009birefringent}. They determine poling periods required for QPM, as summarized in Table~\ref{tab:BMFproperties} together with the second-order nonlinear coefficients \cite{villora2009birefringent,bergman1975linear, chen2012measurement}.
\begin{table}
\centering
    \begin{tabular}{cccc}
\hline\noalign{\smallskip}
Nonlinear \;  & Value \; & Polarizations \; & Poling period  \\
coefficient \; & [pm/V] \; &of photons \; & [nm]\\
\noalign{\smallskip}\hline\noalign{\smallskip}
$d_{32}$ & $0.039$ & bbc & 839 \\
$d_{31}$ & $0.021$ & aac & 984 \\
$d_{33}$ & $0.015$ & ccc & 939 \\
\noalign{\smallskip}\hline
\end{tabular}
\caption{\textbf{Relevant nonlinear coefficient tensor elements} of BMF for SHG at a fundamental wavelength of 1064~ nm, as listed in Ref.~\cite{villora2009birefringent}. Photon polarizations are given in terms of the crystal axes directions for the two incoming photons (first two letters), as well as the outgoing photon (last letter). The last row lists the poling period that would be required for first order QPM to 148.4 nm, assuming Sellmeier parameters listed in Ref.~\cite{villora2009birefringent}.}
\label{tab:BMFproperties}
\end{table}

\section{Periodic poling}
Periodic poling of polished BMF crystals was carried out  at the Fraunhofer-Institut für Physikalische Messtechnik (IPM, Freiburg) adapting a calligraphic poling technology \cite{herr2023fanout}, which allows \textmu m-sized domains in established ferroelectric materials like MgO-doped LiNbO\textsubscript{3}. A \textmu m-sized, electrically-biased tip is moved over the +$c$-face of the BMF crystal to write inverted domains into the bulk crystal defined by the trajectory of the tip. A 50-nm-thick sputtered metal layer (Au) on the -$c$-face of the crystal serves as the counter electrode. Poling is done at room temperature.
By measuring the poling current $I_{p}$, which is given by $2 P_{\text{s}} A/t$, where $P_{\text{s}}$ is the spontaneous polarization of BMF and $A$ the area of the inverted domains within a time $t$, we can accurately monitor the progress of the poling. By controlling the electric bias of the tip via a PID-feedback loop, we keep the poling current constant at a predefined value, which leads, at a given writing speed $v$ (set to 125 \textmu m/s), to a constant, predefined domain width $w$. 
With this, inhomogeneities in the coercive field, for example due to an inhomogeneous stoichiometry or an inhomogeneous scatter center density, can be partly compensated.

The natural choice would be to make use of the largest nonlinear coefficient $d_{32}$, which requires domain lines parallel to the crystallographic $b$-axis. However, preparatory experiments at longer wavelengths (second-harmonic generation from 794- to 397-nm-light via $d_{32}$) have indicated that the temperature-induced change of the refractive indices would not allow for a temperature-based fine-tuning of the QPM condition, which is unfavorable. The reason is that the $dn/dT$-values relevant for $d_{32}$ are compensating each other, leading effectively to no temperature tuning. In addition, the BMF crystals show a tendency of cleaving on the (010) plane, parallel to the crystallographic $a$-axis. For this reason, optical polishing of the fragile crystal facet used for laser input can be avoided by using a cleaved $b$-facet for laser coupling. Furthermore, the achievable minimum domain sizes parallel to the $a$-axis turned out to be narrower than for $b$-parallel domains \cite{herr2023fanout}. For these reasons, we restricted our efforts to the use of the $d_{31}$ and the $d_{33}$ nonlinear coefficients. Notably, both options require the same poling direction with domain lines parallel to the $a$-axis, only the polarization direction of the fundamental laser light needs to be rotated.

For first-order QPM, we would require sub-micron poling periods (cf. Table~\ref{tab:BMFproperties}), which could not be achieved so far. Thus, we need to rely on $m$\textsuperscript{th} order QPM, where the poling period linearly scales with the factor $m$. However, at the same time the effective nonlinear coefficient $d_{\text{eff}}$ reduces by a factor of $m$, leading to a reduction in SHG output by a factor of $m^2$. The achieved minimum poling period with a reasonable quality over the entire crystal length with a 300-\textmu m-thick crystal was 8.86 \textmu m (see Figure \ref{fig:01_BMF}b), which corresponds to 9\textsuperscript{th} order periodic poling using the nonlinear coefficient $d_{31}$.
The long-range order (i.e. constant poling period over the entire crystal) of the 8.86-\textmu m-poling is well preserved. However, variations in the domain width as well as position uncertainties of $\pm 200$ nm of the needle tip lead to random duty-cycle variations, which can lead to further reduction of $d_{\text{eff}}$.

Smaller poling periods of 6.6 \textmu m were reported in \cite{villora2009birefringent}, however, despite significant efforts, we were so far unable to reproduce these results.
When writing domains with a target domain width smaller than about 3-4 \textmu m in 300-\textmu m-thick BMF crystals, we observe a poling current, which proves that poling through the full crystal has occurred. However, after domain-selective etching in nitric acid no inverted domains are observable, which is an indication for unstable domain inversion. Unstable domains in BMF were also identified in prior studies \cite{Mateos2014}.

\section{Optical Setup}
\begin{figure*}
    \centering
    \includegraphics[width=0.8\textwidth]{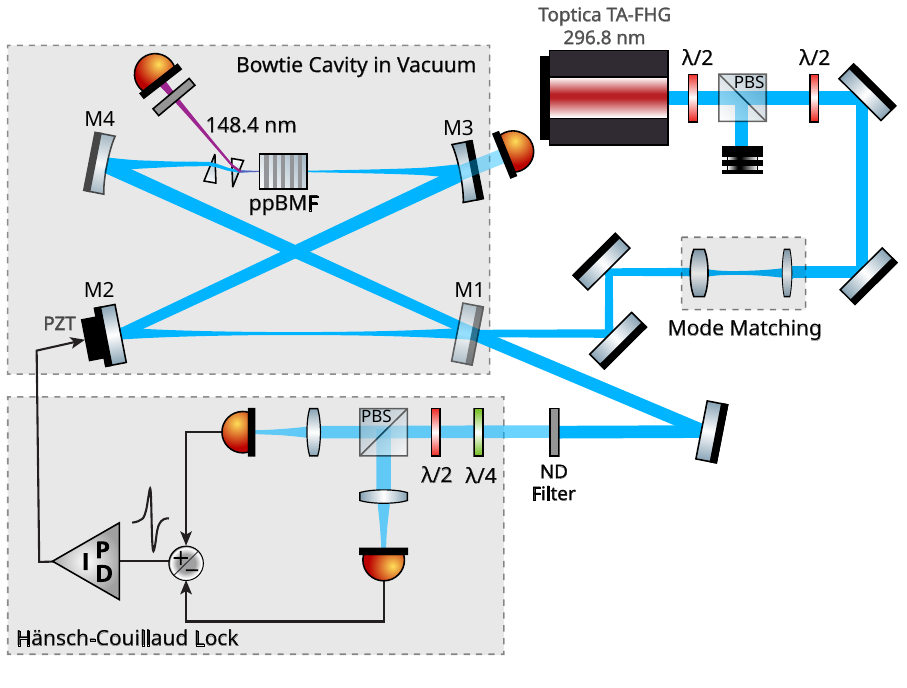}
    \caption{\textbf{Overview of the frequency doubling setup}. Fundamental light at 296.8 nm for SHG is obtained from a commercial TOPTICA TA-FHG pro laser. A combination of a half-wave plate ($\lambda/2$), a Wollaston Prism (PBS) and a half-wave plate ($\lambda/2$) is used to get the desired polarization for the SHG process. A combination of lenses is used to match the transverse mode of the laser beam to the cavity mode. A bowtie cavity consisting of two spherical concave mirrors of radius of curvature of {R=100~mm} leads to a focal beam-waist radius of 30 \textmu m. A periodically poled and temperature controlled BMF (PPBMF) crystal is placed at the focal spot inside the cavity. Two MgF\textsubscript{2} prisms are used for separating the VUV beam from the fundamental recirculating beam. The reflected signal from the cavity is used to actively stabilize its length using a Hänsch-Couillaud lock.} 
    \label{fig:02_Cavity}
\end{figure*}

The optical setup used for cavity-enhanced SHG is shown in Fig.~\ref{fig:02_Cavity} and will be described in detail in the following. Fundamental continuous-wave laser light at 296.8~nm is generated using a commercial \mbox{TOPTICA TA-FHG pro} laser which consists internally of two frequency doubling stages 1187.2~nm $\rightarrow$ 593.6~nm $\rightarrow$ 296.8~nm. This laser system delivers 600 mW fundamental light on a continuous basis and up to 800 mW short time. This fundamental laser leads to a SHG center frequency (wavelength) of 2020.4~THz (148.4~nm) which can be tuned by $\pm400$~GHz. For a given fundamental power, the VUV power is constant over this frequency range. In addition, with the fundamental laser free-running a linewidth of 480~kHz at 148.4~nm is obtained.

To match the Brewster condition of the outcoupling prisms inside the enhancement resonator the fundamental light should be p-polarized. A half wave plate ($\lambda/2$), polarizing beam splitter (PBS) and another half wave plate are used to prepare the required polarization. Further, a two-lens telescope is used to match the transverse mode of the fundamental laser with that of the enhancement cavity. The fundamental light is then coupled to a UV enhancement cavity designed and built by Agile Optic GmbH. Since the generated VUV light is strongly absorbed in air, the cavity is designed to be vacuum compatible and placed in a vacuum chamber at a pressure of about $10^{-6}$~mbar. To avoid degradation of the optical elements due to cracking of hydrocarbon molecules from irradiation with intense 296.8~nm light and the resulting 148.4~nm radiation (hydrocarbon contamination), the vacuum chamber is purged with an ozone-oxygen mixture ($3.5\%$ Ozone by weight) at a background pressure of about 5x10\textsuperscript{-3}~mbar during operation \cite{vig1985uv}. For this purpose, an ozone generator (Oxidation Technologies, Type Enaly 1KNT) is connected to an oxygen gas bottle and the generated ozone-oxygen mixture is injected into the vacuum chamber with a leak-valve.  Ozone, being a highly reactive oxidizing agent, is used for removal of the carbon layer due to the formation of CO and CO\textsubscript{2}. Notably at these ozone-oxygen concentrations and background pressures, residual-gas absorption of 148.4 nm radiation over 10~cm remains negligible at $1.8\%$. In addition, this constant purging is ultimately indispensable for continuous, long-term operation without a reduction in VUV power over time.

For the enhancement resonator a bowtie configuration is chosen, because it prevents the formation of a standing-wave inside the cavity which would otherwise lead to regions in the crystal with alternating high and low intensities. A linear standing-wave geometry would additionally result in two lower-power SHG output beams instead of a single one. Fundamental light is coupled into the cavity using a plane incoupling mirror (M1) with a reflectivity of 75\%. The beam is then incident on a mirror mounted on a piezo actuator (M2) which is used to keep the cavity resonant with the fundamental laser. Two concave mirrors of radius of curvature $R = 100$~mm (M3 and M4) are used to focus the fundamental light onto the nonlinear crystal. The uncoated PPBMF crystal of dimensions (a $\times$ b $\times$ c) $5\times5\times 0.3$ mm\textsuperscript{3} is mounted under normal incidence on a translation stage centered between the two curved mirrors. The crystal mount is actively temperature-stabilized allowing for a crystal temperature range of $16\degree$C to $80\degree$C. Changing the temperature of the crystal predominantly modifies the refractive indices thereby slightly changing the coherence length. When the coherence length matches the size of each domain, the quasi-phase matching condition is satisfied and results in a larger VUV signal. A first VUV-grade MgF\textsubscript{2} prism (Korth Kristalle) at Brewster angle for the fundamental light is used for outcoupling of the SHG light from the cavity. This avoids cutting the SHG crystal to Brewster angle, thereby avoiding any crystal-specific cavity design and providing maximum flexibility to use different types of nonlinear crystals. A second MgF\textsubscript{2} prism redirects the fundamental beam onto the second concave mirror M4, thereby closing the cavity. The resulting cavity length (L) of 600 mm corresponds to a free spectral range (FSR) of $500$~MHz. The combination of the mirror curvatures and the cavity length leads to cavity parameters $g_1 = -4.1$ and $g_2 = -0.1$, well within the stability regions of cavity operation \cite{kogelnik1966laser}. The described geometry and parameters correspond to cavity transverse modes with a focal beam-waist radius of 30 \textmu m at the crystal and 127  \textmu m (horizontal), 153 \textmu m (vertical) between the incoupling (M1) and piezo mirror (M2). The outcoupling prims currently limit how close the curved mirrors can be to the crystal thereby determining their smallest possible radius of curvatures and focal spot sizes at the crystal.

The cavity length is stabilized using a Hänsch-Couillaud \cite{hansch1980laser} lock to enable the build-up of intracavity fundamental power. The Brewster-cut MgF\textsubscript{2} prisms act as a polarizing element blocking the build-up of s-polarized light in the cavity while the p-polarized light accumulates a cavity length dependent phase shift. This leads to a change in polarization ellipticity of the beam on the cavity reflection path which is measured using a combination of a quarter wave plate ($\lambda/4$), half wave plate ($\lambda/2$) and a PBS. A differential measurement on the two ports of the PBS gives a dispersive error signal which is then used to stabilize the cavity length as shown in Figure \ref{fig:02_Cavity}.

Two different detector systems are used to characterize the VUV light outcoupled from the cavity:

(1) a microchannel plate (MCP) detector with a phosphor screen for spatially resolved photon detection (GIDS, type MCP-25-2-40-P43), mounted at a distance of approximately 140 mm from the BMF crystal. The phosphor screen is monitored with a CCD camera located outside of the vacuum chamber. The MCP front surface is coated with a 50~nm thin layer of CsI to enhance the VUV detection efficiency. Since the fundamental power is large compared to the SHG power, two VUV bandpass filters (Acton  FN147-N-1D) are used in front of the detector to attenuate the fundamental light. The transmission of each filter at 296.8~nm and 148.4~nm is about $0.025\%$ and $21\%$ respectively. This additional attenuation is required, even though CsI has a very low detection efficiency of potentially below 10\textsuperscript{-7}\% for 296.8~nm light \cite{hamamatsu_r1081datasheet}. Overall the 296.8~nm background is below $1\%$ for the largest measured powers and no background subtraction is performed in the data.

(2) A photomultiplier tube (PMT, Hamamatsu type R1081, CsI photocathode) operated in single-photon counting mode in combination with a photon counter (Vertilon, type PhotonIQ MCPC618).  As with the MCP, two VUV bandpass filters are used in front of the PMT to attenuate the fundamental light.

\section{Measurements}
\begin{figure}
    \centering
    \includegraphics[width=0.95\linewidth]{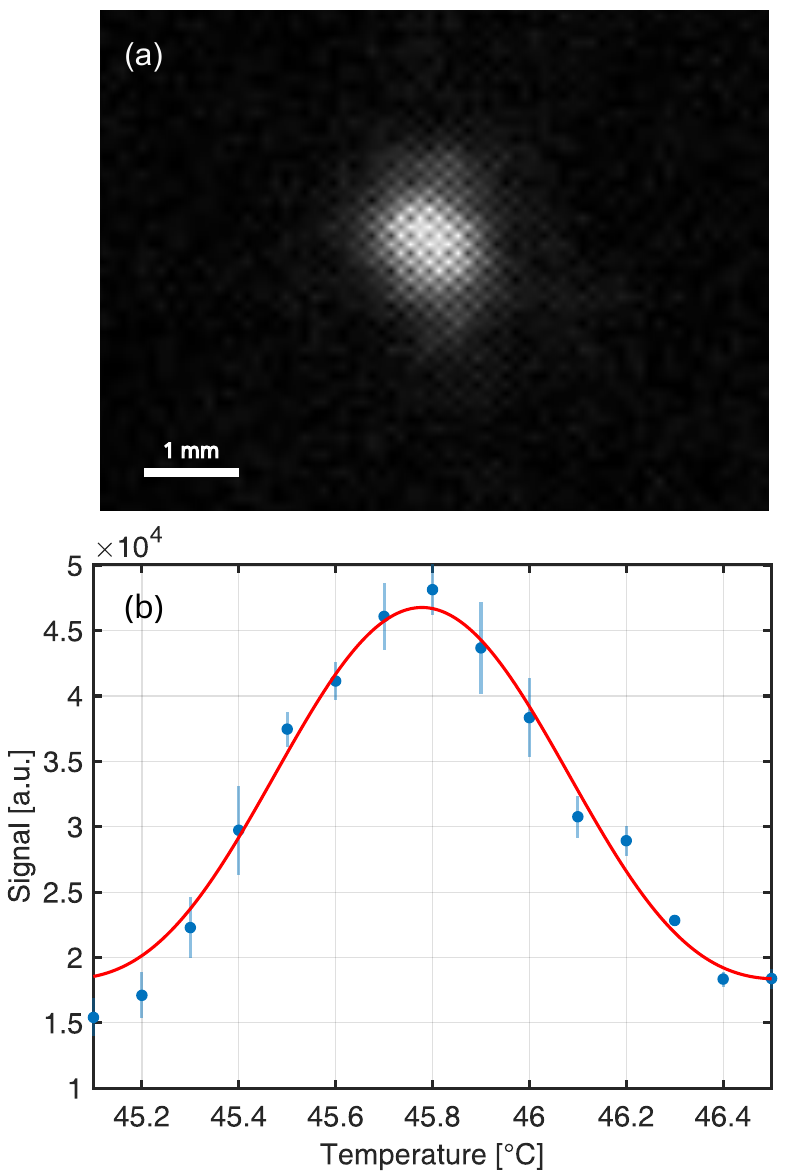}
    \caption{\textbf{Second harmonic generation using PPBMF.} For a first detection of the SHG light at 148.4 nm, a CsI-coated microchannel plate (MCP) with a phosphor screen is used. Two optical filters are used in front of the MCP to filter out any stray fundamental light. (a) Image of the MCP phosphor screen showing the VUV beam. The periodic grid seen reflects the granularity of the MCP detector due to the micro-channels. (b) The strength of the VUV beam is strongly dependent on temperature and is used for fine-tuning the quasi-phase-matching condition. The blue data points are the signal obtained by summing counts from a region of interest (ROI) on the image of the MCP phosphor screen, while the red line is a fit to a sinc\textsuperscript{2} function plus offset.}
    \label{fig:03_SHG}
\end{figure}

Measurement of the intracavity power was performed with the transmission photo-diode after calibration. Without the BMF crystal, a circulating power of about 1.8 W at 400 mW incoming power was measured. This corresponds to an enhancement factor of about $4.5$ and a cavity finesse of about 11.6, limited by the reflectivity of the input coupler ($\sim75\%$), as well as roundtrip losses in the MgF$_2$ prisms ($\sim20$\%). When the BMF crystal was mounted, the maximum intracavity power dropped to about 1.4 W at 600 mW input power, corresponding to an enhancement factor of about 2.3 and a finesse of 7.8. This reduction can be attributed to additional roundtrip losses of $\sim20$\% due to the uncoated BMF crystal.

The MCP detector was used for a first detection as well as a later alignment of the VUV beam. An image of the resulting VUV radiation on the MCP is shown in Figure \ref{fig:03_SHG}a. The $1/e^2$ radius of the spot is approximately 300~\textmu m, corresponding to a divergence half-angle of the SHG light of approximately $0.13\degree$. Figure \ref{fig:03_SHG}b shows the VUV signal as a function of the crystal temperature. Close to the maximum, this signal resembles a sinc$^2$ curve plus offset as a function of the phase slip \cite{boyd2008nonlinear}. Even though the poling dimensions are tailored for polarizations of the fundamental and VUV corresponding to $d_{31}$, an about 10x stronger signal is observed for polarizations corresponding to $d_{33}$. All provided data is therefore based on the $d_{33}$ nonlinear coefficient.

\begin{figure}
    \centering
    \includegraphics[width=0.95\linewidth]{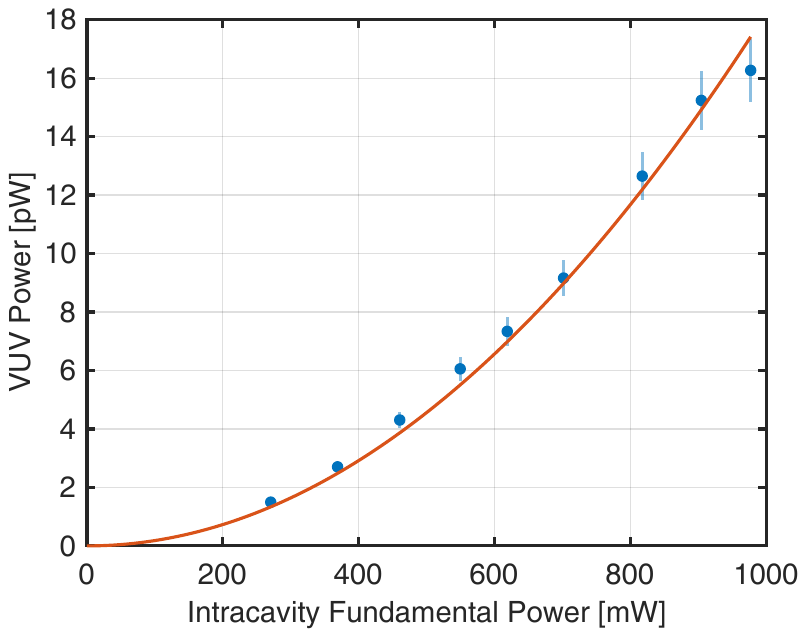}
    \caption{\textbf{VUV beam power scaling.} Using a Photo-Multiplier Tube (PMT) the power in the VUV beam is measured. The VUV power $P_2$ scales as the square of the fundamental power $P_{1}^2$ as expected (red line) for second harmonic generation. The error bars include systematic and measurement uncertainty.}
    \label{fig:04_Power_Scaling}
\end{figure}

\begin{figure}
    \centering
    \includegraphics[width=0.95\linewidth]{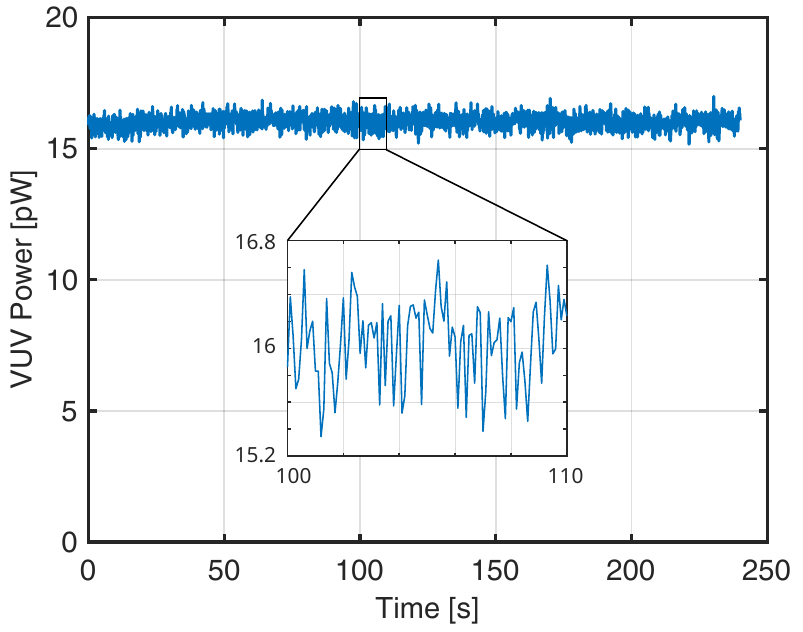}
    \caption{\textbf{Stability of the VUV laser.} VUV power is monitored on the PMT with the cavity locked and temperature of the crystal fixed. The power remains stable when the vacuum chamber, in which the cavity is placed, is purged continuously with and ozone-oxygen mixture at a pressure of about 5x10\textsuperscript{-3}~mbar. In the absence of purging both the fundamental and VUV powers drop slowly due to possible carbon layer build-up on the optical surfaces.}
    \label{fig:05_VUV_Timeseries}
\end{figure}

In order to quantify the VUV power, the MCP was replaced with the PMT. From the known PMT quantum efficiency ($9.34 \pm 0.70~\%$)  and the transmission characteristics of the VUV bandpass filters ($21.0\pm1.6~\%$ per filter), the power of the VUV radiation is extracted. This is shown in Figure \ref{fig:04_Power_Scaling} as a function of the intracavity fundamental power and shows a quadratic scaling characteristic of a $\chi^{(2)}$-nonlinear frequency doubling process, as expected in the undepleted pump regime. The typically achieved SHG power level is $(16\pm1)$~pW, although a position-dependent maximum of $(48\pm3)$~pW at 1.4~W intra-cavity fundamental power was observed. Notably, some significant crystal-position dependence of the obtained SHG power was observed. This is understandable in view of the present duty-cycle variations (see Figure \ref{fig:01_BMF}b), which are expected to have a significant influence on the generated SHG power at 9\textsuperscript{th} order QPM. The reason is, that only 1/9th of the domain width is effectively used for SHG build-up, corresponding to about 492~nm, while variation of the domain width is larger. Furthermore, the crystal transparency is observed to be position dependent resulting in variations of cavity-enhanced fundamental power. To demonstrate the stability of the VUV laser, a time series of the VUV power is shown in Figure~\ref{fig:05_VUV_Timeseries}. The cavity is locked and the SHG power is stable for more than 200~s. When the vacuum chamber is not purged with ozone, the intracavity fundamental as well as VUV powers drop on a time scale of a few tens of seconds. This process is found to be completely reversible with ozone purging.

\section{Discussion}
In this work intra-cavity SHG to 148.4~nm based on periodically poled BMF crystals was experimentally achieved for the first time, aiming at the development of a novel type of CW laser for nuclear optical clock operation. The maximum observed SHG power of $(48\pm3)$~pW at 1.4~W intra-cavity fundamental power compares to a theoretically expected value of $(63\pm18)$~pW. The theoretical estimate is based on the Boyd-Kleinman equation \cite{boyd1968parametric,daniel2020analytical,sah2025master}
$$
P_\text{2}=\frac{2T\omega_1^2 d_\text{eff}^2 L k_1 P_1^2 h}{\pi\epsilon_0 c^3 n_1^2 n_2},
$$
with power of frequency doubled light $P_2$, crystal transmission $T$, angular frequency of the fundamental light $\omega_1$, effective nonlinear coefficient $d_\text{eff}=2/(m\pi)\cdot d$ with poling order $m$, nonlinear coefficient $d$, crystal length $L$, wave number of the fundamental light $k_1$, power of the fundamental light $P_1$, vacuum permittivity $\epsilon_0$, speed of light $c$, index of refraction of the fundamental light $n_1$ and of the SHG light $n_2$. Further, $h$ is the value of the Boyd-Kleinman integral \cite{daniel2020analytical}. To model our experimental conditions, we used parameters of $T=3.6\%$, $d=0.015$ pm/V, $m=9$, $L=5$ mm, 30 \textmu m beam waist radius and 1.4~W cavity-enhanced fundamental power $P_1$. Notably, the theoretical estimate is only by a factor of 1.3 larger than the experimentally obtained maximum value.

Under more favorable, but realistic conditions (e.g., 50\% VUV crystal transmission, 15 \textmu m beam waist radius, 10~W cavity-enhanced fundamental power, while keeping all remaining parameters constant) the calculated SHG power scales up to 156 nW. When further assuming that an improved poling order and a larger nonlinear coefficient could be made use of (e.g., $m=3$ and $d=0.039$ pm/V), the theoretically obtained SHG power rises to 9.5~\textmu W. Therefore, we assume that significant improvements of BMF-based output powers will be obtained in the future. These estimates are fully compatible with the laser induced damage threshold of BMF which is expected to be atleast 1~GW/cm\textsuperscript{2}\cite{bergman1975linear}.

For increased VUV powers, we are planning to implement the following improvements: Firstly, a higher cavity finesse will be reached by replacing the two MgF\textsubscript{2} prisms with a dichroic optical element and using a different cavity in-coupler as well as anti-reflection coated crystals. In addition, a smaller radius-of-curvature of the cavity mirrors will lead to a smaller beam waist. Secondly, using a crystal of higher VUV transparency is expected to result in a significantly enhanced output power. This may, however, come at the cost of poling quality. 
Thirdly, the observed $d_{31}-d_{33}$ anomaly points towards non-optimal poling dimensions and a required correction of the Sellmeier coefficients. A re-evaluation of Sellmeier coefficients based on refractive index measurements in the UV/VUV range is underway.
  
In addition, work is ongoing to obtain smaller poling periods, thereby allowing lower QPM orders in the future - for example going to 3\textsuperscript{rd} instead of 9\textsuperscript{th} order could improve the VUV yield by a factor of 9 or beyond, if the variation of the domain width is improved. To this end, we aim to use thinner BMF crystals as this is expected to allow considerably smaller stable domains \cite{Mateos2014, kan2007}. Ultimately, we are striving toward \textmu m-sized ridge waveguides, as this will dramatically reduce the required crystal thickness, bringing even 1\textsuperscript{st} order QPM into reach \cite{Nagy:20}. At the same time the strong light confinement, which comes with a \textmu m-sized waveguide, can boost the nonlinear light-matter interaction, compensating for the intrinsically low nonlinear coefficients of BMF. As an alternative for traditional QPM via electric-field poling, the method of order-disorder phase matching via direct laser lithography as discussed in Ref. \cite{shao2020pushing} could also be applied to BMF and is subject to ongoing work as well. As this method is not restricted to ferroelectric crystals, it might be also an attractive prospect for SBO and BPO$_4$ which offer larger nonlinear coefficients.
Finally, it should be pointed out that the recent advances in growing periodically patterned SBO crystals \cite{perlov2024method} is expected to lead to drastically improved SHG powers at 148.4 nm in the near future, because SBO offers nonlinear coefficients in the range of 1.5~pm/V, a factor of about 100 larger compared to the $\chi^{(2)}$-value utilized in this work. With the same experimental setup, one would expect four orders of magnitude larger SHG power when using periodically patterned SBO instead of PPBMF.

\begin{acknowledgments}
We acknowledge discussions with S.~Stellmer, J.~Ye,  E.~R.~Hudson, T.~Schumm, T.~Udem, C.~Zhang, T.~Ooi, J.~F.~Doyle, K.~Li, Ch.~E.~Düllmann, J.~Stricker, E.~G.~Villora, K.~Shimamura and V.~Ya.~Shur. This work is supported by German Federal Ministry of Research, Technology and Space (BMFTR) Quantum Futur II Grant ``NuQuant'' under grant number 13N16295A, UV-KrisP under grant number 13N16924, the Cluster of Excellence  ``Precision Physics, Fundamental Interactions, and Structure of Matter'' (PRISMA++ EXC 2118/2) funded by the German Research Foundation (DFG) within the German Excellence Strategy (Project ID 390831469) and Quantum Futur II Grant "SemIQON"
under grant number 13N16291.
\end{acknowledgments}
\bibliography{Bibliography}

@PREAMBLE{
 "\providecommand{\noopsort}[1]{}" 
 # "\providecommand{\singleletter}[1]{#1}%" 
}

@article{dimarcq2024roadmap,
  title={Roadmap towards the redefinition of the second},
  author={Dimarcq, Noel and Gertsvolf, Marina and Mileti, Gaetano and Bize, Sebastien and Oates, CW and Peik, Ekkehard and Calonico, Davide and Ido, Tetsuya and Tavella, Patrizia and Meynadier, Fr{\'e}d{\'e}ric and others},
  journal={Metrologia},
  volume={61},
  number={1},
  pages={012001},
  year={2024},
  publisher={IOP Publishing},
  url={https://doi.org/10.1088/1681-7575/ad17d2}
}

@article{ludlow2015optical,
  title={Optical atomic clocks},
  author={Ludlow, Andrew D and Boyd, Martin M and Ye, Jun and Peik, Ekkehard and Schmidt, Piet O},
  journal={Reviews of Modern Physics},
  volume={87},
  number={2},
  pages={637--701},
  year={2015},
  publisher={APS},
  url={https://doi.org/10.1103/RevModPhys.87.637}
}

@article{kozlov2018highly,
  title={Highly charged ions: Optical clocks and applications in fundamental physics},
  author={Kozlov, MG and Safronova, MS and Crespo L{\'o}pez-Urrutia, JR and Schmidt, PO},
  journal={Reviews of Modern Physics},
  volume={90},
  number={4},
  pages={045005},
  year={2018},
  publisher={APS},
  url={https://doi.org/10.1103/RevModPhys.90.045005}
}

@article{peik2003nuclear,
  title={Nuclear laser spectroscopy of the 3.5 {eV} transition in {Th-229}},
  author={Peik, E and Tamm, Chr},
  journal={Europhysics Letters},
  volume={61},
  number={2},
  pages={181},
  year={2003},
  publisher={IOP Publishing},
  url={https://doi.org/10.1209/epl/i2003-00210-x}
}

@article{von2016direct,
  title={Direct detection of the {$^{229}$Th} nuclear clock transition},
  author={von der Wense, Lars and Seiferle, Benedict and Laatiaoui, Mustapha and Neumayr, J{\"u}rgen B and Maier, Hans-J{\"o}rg and Wirth, Hans-Friedrich and Mokry, Christoph and Runke, J{\"o}rg and Eberhardt, Klaus and D{\"u}llmann, Christoph E and others},
  journal={Nature},
  volume={533},
  number={7601},
  pages={47--51},
  year={2016},
  publisher={Nature Publishing Group UK London},
  url={https://doi.org/10.1038/nature17669}
}

@article{rellergert2010constraining,
  title={Constraining the Evolution of the Fundamental Constants with a Solid-State Optical Frequency Reference Based on the {$^{229}$Th} Nucleus},
  author={Rellergert, Wade G and DeMille, D and Greco, Richard R and Hehlen, Markus Peter and Torgerson, JR and Hudson, Eric R},
  journal={Physical review letters},
  volume={104},
  number={20},
  pages={200802},
  year={2010},
  publisher={APS},
  url={https://doi.org/10.1103/PhysRevLett.104.200802}
}

@article{vonderwense2020review,
  title={The {$^{229}$Th} isomer: prospects for a nuclear optical clock},
  author={von der Wense, Lars and Seiferle, Benedict},
  journal={Europ. Phys. J. A},
  volume={56},
  pages={277},
  year={2020},
  url={https://doi.org/10.1140/epja/s10050-020-00263-0}
}

@article{tiedau2024laser,
  title={Laser excitation of the {Th-229} nucleus},
  author={Tiedau, J and Okhapkin, MV and Zhang, K and Thielking, J and Zitzer, G and Peik, E and Schaden, Fabian and Pronebner, T and Morawetz, Ira and De Col, L Toscani and others},
  journal={Physical Review Letters},
  volume={132},
  number={18},
  pages={182501},
  year={2024},
  publisher={APS},
  url={https://doi.org/10.1103/PhysRevLett.132.182501}
}

@article{elwell2024laser,
  title={Laser excitation of the {$^{229}$Th} nuclear isomeric transition in a solid-state host},
  author={Elwell, R and Schneider, Christian and Jeet, Justin and Terhune, JES and Morgan, HWT and Alexandrova, AN and Tran Tan, HB and Derevianko, Andrei and Hudson, Eric R},
  journal={Physical Review Letters},
  volume={133},
  number={1},
  pages={013201},
  year={2024},
  publisher={APS},
  url={https://doi.org/10.1103/PhysRevLett.133.013201}
}

@article{zhang2024frequency,
  title={Frequency ratio of the {$^{229\text{m}}$Th} nuclear isomeric transition and the {$^{87}$Sr} atomic clock},
  author={Zhang, Chuankun and Ooi, Tian and Higgins, Jacob S and Doyle, Jack F and von der Wense, Lars and Beeks, Kjeld and Leitner, Adrian and Kazakov, Georgy A and Li, Peng and Thirolf, Peter G and others},
  journal={Nature},
  volume={633},
  number={8028},
  pages={63--70},
  year={2024},
  publisher={Nature Publishing Group UK London},
  url={https://doi.org/10.1038/s41586-024-07839-6}
}

@article{lal2025continuous,
  title={Continuous-wave laser source at the 148 nm nuclear transition of {Th-229}},
  author={Lal, Vishal and Okhapkin, Maksim V and Tiedau, Johannes and Irwin, Niels and Petrov, Valentin and Peik, Ekkehard},
  journal={Optica},
  volume={12},
  number={12},
  pages={1971--1974},
  year={2025},
  publisher={Optica Publishing Group},
  url={https://doi.org/10.1364/OPTICA.574489}
}

@article{morawetz2026cwspectroscopy,
  title={Continuous-wave nuclear laser absorption spectroscopy of {Thorium-229}},
  author={I. Morawetz and T. Riebner and L. Toscani De Col and F. Schneider and N. Sempelmann and F. Schaden and M. Bartokos and G. A. Kazakov and S. Lahs and K. Beeks and others},
  journal={arXiv},
  pages={2604.16640},
  year={2026},
  publisher={arXiv Preprint},
  url={https://doi.org/10.48550/arXiv.2604.16640}
}

@article{Toscani2026,
  title={A thorium-229 optical nuclear clock with feedback loop},
  author={L. Toscani De Col and T. Riebner and I. Morawetz and F. Schneider and N. Sempelmann  and J. Schlachet-L´epinay and F. Schaden and M. Bartokos and G. A. Kazakov and K. Beeks and B. Gerstenecker and M. Pimon and S. Lahs and
A. Hellerschmied and T. Lercher and J. Premper and A. Niessner and M. Matus and H. Denker and M. Cizek and O. Cip and V. Lal and G. Zitzer and V. Petrov and J. Tiedau and M. V. Okhapkin and E. Peik and T. Schumm},
  journal={arXiv},
  pages={2606.04997},
  year={2026},
  publisher={arXiv Preprint},
  url={https://doi.org/10.48550/arXiv.2606.04997}
}

@article{xiao2026continuous,
  title={Continuous-wave narrow-linewidth vacuum ultraviolet laser source},
  author={Xiao, Qi and Penyazkov, Gleb and Li, Xiangliang and Huang, Beichen and Bu, Wenhao and Shi, Juanlang and Shi, Haoyu and Liao, Tangyin and Yan, Gaowei and Tian, Haochen and others},
  journal={Nature},
  pages={1--5},
  year={2026},
  publisher={Nature Publishing Group UK London},
  url={https://doi.org/10.1038/s41586-026-10107-4}
}

@article{Huang2026,
  title={A nuclear clock based on {$^{229}$Th}},
  author={Beichen Huang and Gaowei Yan and Qi Xiao and Wenhao Bu and Zhen Zhang and Chengchun Zhao and Chao Yan and Zhi-Ang Chen and Peixiong Zhang and Gleb Penyazkov and Zhenhai Zhan and Lingfeng
Yan and Yuefei Wang and Lin Li and Shanming Li and Xiaobo Qian and Xuegang Liu and Qiange He and Taoxiang Sun and
Haochen Tian and Binkun Lu and Ningyuan Ma and Juxian Li and Yanzhang Wu and Qiaorui Gong and Yuxiang Li and Haoyu Shi and Xiangliang Li and Longsheng Ma and Shining Zhu and Yuxiang Mo and Jun Lin and Li
You and Yige Lin and Xibo Zhang and Yin Hang and Liangbi Su and Shiqian Ding},
  journal={arXiv},
  pages={2606.08870},
  year={2026},
  publisher={arXiv Preprint},
  url={https://doi.org/10.48550/arXiv.2606.08870}
}

@article{shao2020pushing,
  title={Pushing periodic-disorder-induced phase matching into the deep-ultraviolet spectral region: theory and demonstration},
  author={Shao, Mingchuan and Liang, Fei and Yu, Haohai and Zhang, Huaijin},
  journal={Light: Science \& Applications},
  volume={9},
  number={1},
  pages={45},
  year={2020},
  publisher={Nature Publishing Group UK London},
  url={https://doi.org/10.1038/s41377-020-0281-4}
}

@article{buchter2001periodically,
  title={Periodically poled {BaMgF\textsubscript{4}} for ultraviolet frequency generation},
  author={Buchter, SC and Fan, TY and Liberman, V and Zayhowski, JJ and Rothschild, M and Mason, EJ and Cassanho, A and Jenssen, HP and Burnett, John H},
  journal={Optics Letters},
  volume={26},
  number={21},
  pages={1693--1695},
  year={2001},
  publisher={Optical Society of America},
  url={https://doi.org/10.1364/OL.26.001693}
}

@article{villora2009birefringent,
  title={Birefringent-and quasi phase-matching with {BaMgF$_4$} for vacuum-{UV/UV} and {mid-IR} all solid-state lasers},
  author={V{\'\i}llora, Encarnaci{\'o}n G and Shimamura, Kiyoshi and Sumiya, Keiji and Ishibashi, Hiroyuki},
  journal={Optics express},
  volume={17},
  number={15},
  pages={12362--12378},
  year={2009},
  publisher={Optical Society of America},
  url={https://doi.org/10.1364/OE.17.012362}
}

@article{zaitsev2008domain,
  title={Domain structure in strontium tetraborate single crystal},
  author={Zaitsev, AI and Aleksandrovsky, AS and Vasiliev, AD and Zamkov, AV},
  journal={Journal of crystal growth},
  volume={310},
  number={1},
  pages={1--4},
  year={2008},
  publisher={Elsevier},
  url={https://doi.org/10.1016/j.jcrysgro.2007.09.037}
}

@misc{perlov2024method,
  title={Method for manufacturing of patterned {SrB4BO7} and {PbB4O7} crystals},
  author={Perlov, Dan and Zaytsev, Alexander and Zamkov, Anatolii and Radinov, Nikita and Cherepakhin, Aleksandr and Evtikhiev, Nikolay and Sadovskiy, Andrey},
  year={2022},
  month=jan # "~9",
  publisher={Google Patents},
  note={{US} Patent 11,868,022},
  url={https://patents.google.com/patent/US11868022B2/en}
}

@inproceedings{vasilyev2026148nm,
  title={148 nm vacuum ultraviolet laser source for high-resolution spectroscopy},
  author={Vasilyev, Sergey and Moskalev, Igor and Ooi, Tian and Mirov, Mike and Muraviev, Andrey and Konnov, Dmitrii and Churikov, Victor and Sukharev, Viktor and Galenin, Evgeny and Doyle, John and others},
  booktitle={Nonlinear Frequency Generation and Conversion: Materials and Devices XXV},
  volume={13877},
  pages={1387706},
  year={2026},
  organization={SPIE},
  url={https://doi.org/10.1117/12.3081827}
}

@article{morgan2025proposal,
  title={Proposal and theoretical investigation of {$^{229}$Th}-doped nonlinear optical crystals for compact solid-state clocks},
  author={Morgan, H\_W T and Elwell, R and Terhune, JES and Tran Tan, HB and Perera, UC and Derevianko, A and Alexandrova, AN and Hudson, ER},
  journal={Applied Physics Letters},
  volume={126},
  number={11},
  year={2025},
  publisher={AIP Publishing},
  url={https://doi.org/10.1063/5.0247867}
}

@article{herr2023fanout,
  title={Fanout periodic poling of {BaMgF\textsubscript{4}} crystals},
  author={Herr, Simon J and Tanaka, Hiroki and Breunig, Ingo and Bickermann, Matthias and K{\"u}hnemann, Frank},
  journal={Optical Materials Express},
  volume={13},
  number={8},
  pages={2158--2164},
  year={2023},
  publisher={Optica Publishing Group},
  url={https://doi.org/10.1364/OME.492170}
}

@article{zhao2011formation,
  title={Formation mechanism of scattering centers in {BaMgF\textsubscript{4}} single crystals},
  author={Zhao, Chengchun and Zhang, Lianhan and Hang, Yin and He, Xiaoming and Yin, Jigang and Hu, Pengchao and Chen, Guangzhu and He, Mingzhu and Huang, Huang and Zhu, Yongyuan},
  journal={Journal of Crystal Growth},
  volume={316},
  number={1},
  pages={158--163},
  year={2011},
  publisher={Elsevier},
  url={https://doi.org/10.1016/j.jcrysgro.2010.12.053}
}

@article{bergman1975linear,
  title={Linear and nonlinear optical properties of ferroelectric {BaMgF\textsubscript{4}} and {BaZnF\textsubscript{4}}},
  author={Bergman, JG and Crane, GR and Guggenheim, H},
  journal={Journal of Applied Physics},
  volume={46},
  number={11},
  pages={4645--4646},
  year={1975},
  publisher={American Institute of Physics},
  url={https://doi.org/10.1063/1.321541}
}

@article{chen2012measurement,
  title={Measurement of second-order nonlinear optical coefficients of {BaMgF\textsubscript{4}}},
  author={Chen, Junjie and Chen, Xianfeng and Ma, Yanzhi and Zheng, Yuanlin and Wu, Anhua and Li, Hongjun and Jiang, Linwen and Xu, Jun},
  journal={Journal of the Optical Society of America B},
  volume={29},
  number={4},
  pages={665--668},
  year={2012},
  publisher={Optical Society of America},
  url={https://doi.org/10.1364/JOSAB.16.000620}
}

@article{Mateos2014,
author = {Mateos, Luis and Ramírez, Mariola O. and Carrasco, Irene and Molina, Pablo and Galisteo-López, Juan F. and Víllora, Encarnación G. and de las Heras, Carmen and Shimamura, Kiyoshi and Lopez, Cefe and Bausá, Luisa E.},
title = {{BaMgF\textsubscript{4}}: An Ultra-Transparent Two-Dimensional Nonlinear Photonic Crystal with Strong {\textchi\textsuperscript{(3)}} Response in the UV Spectral Region},
journal = {Advanced Functional Materials},
volume = {24},
number = {11},
pages = {1509-1518},
doi = {https://doi.org/10.1002/adfm.201302588},
url = {https://advanced.onlinelibrary.wiley.com/doi/abs/10.1002/adfm.201302588},
year = {2014}
}

@article{vig1985uv,
  title={UV/ozone cleaning of surfaces},
  author={Vig, John R},
  journal={Journal of Vacuum Science \& Technology A: Vacuum, Surfaces, and Films},
  volume={3},
  number={3},
  pages={1027--1034},
  year={1985},
  publisher={American Vacuum Society},
  url={ https://doi.org/10.1116/1.573115}
}

@article{kogelnik1966laser,
  title={Laser beams and resonators},
  author={Kogelnik, Herwig and Li, Tingye},
  journal={Applied optics},
  volume={5},
  number={10},
  pages={1550--1567},
  year={1966},
  publisher={Optical Society of America},
  url={https://doi.org/10.1364/AO.5.001550}
}

@article{hansch1980laser,
  title={Laser frequency stabilization by polarization spectroscopy of a reflecting reference cavity},
  author={H{\"a}nsch, TW and Couillaud, B},
  journal={Optics communications},
  volume={35},
  number={3},
  pages={441--444},
  year={1980},
  publisher={Elsevier},
  url={https://doi.org/10.1016/0030-4018(80)90069-3}
}

@book{photonics2000photomultiplier,
  title={Photomultiplier tubes},
  author={Photonics, Hamamatsu},
  year={2000},
  publisher={Hamamatsu},
  url={https://www.hamamatsu.com/content/dam/hamamatsu-photonics/sites/documents/99_SALES_LIBRARY/etd/PMT_handbook_v4E.pdf}
}

@manual{hamamatsu_r1081datasheet,
  title        = {Hamamatsu R1081 Photomultiplier Tube Datasheet},
  organization = {Hamamatsu Photonics},
  url          = {https://www.hamamatsu.com/content/dam/hamamatsu-photonics/sites/documents/99_SALES_LIBRARY/etd/R1081_TPMH1119E.pdf}
}

@incollection{boyd2008nonlinear,
  title={Nonlinear optics},
  author={Boyd, Robert W and Gaeta, Alexander L and Giese, Enno},
  booktitle={Springer handbook of atomic, molecular, and optical physics},
  pages={1097--1110},
  year={2008},
  publisher={Springer},
  url={https://doi.org/10.1007/978-3-030-73893-8_76}
}

@article{boyd1968parametric,
  title={Parametric interaction of focused Gaussian light beams},
  author={Boyd, G. D. and Kleinman, D. A. },
  journal={Journal of Applied Physics},
  volume={39},
  number={8},
  pages={3597--3639},
  year={1968},
  url={https://doi.org/10.1063/1.1656831}
}

@article{daniel2020analytical,
  title={Analytical approximation of the second-harmonic conversion efficiency},
  author={Daniel, John R. and Tsai, Shan-Wen and Hemmerling, Boerge},
  journal={Applied Optics},
  volume={59},
  number={28},
  pages={9010--9014},
  year={2020},
}

@article{sah2025master,
    author = {Nutan Kumari Sah},
    title = {Towards a Continuous-Wave {VUV} Laser for
Thorium-229m Nuclear Clock},
    journal = {Master thesis, JGU Mainz},
    year = {2025},
    url={https://doi.org/10.5281/zenodo.20356406},
}

@article{kan2007,
 author = {Kan, Yi and Lu, Xiaomei and Bo, Huifeng and Huang, Fengzhen and Wu, Xiaobo and Zhu, Jinsong},
 year = {2007},
 title = {Critical radii of ferroelectric domains for different decay processes in LiNbO3 crystals},
 pages = {132902},
 volume = {91},
 number = {13},
 issn = {0003-6951},
 journal = {Applied Physics Letters},
 doi = {10.1063/1.2790475}
}

@article{Nagy:20,
author = {Jonathan Tyler Nagy and Ronald M. Reano},
journal = {Opt. Mater. Express},
number = {8},
pages = {1911--1920},
publisher = {Optica Publishing Group},
title = {Submicrometer periodic poling of lithium niobate thin films with bipolar preconditioning pulses},
volume = {10},
month = {Aug},
year = {2020},
url = {https://opg.optica.org/ome/abstract.cfm?URI=ome-10-8-1911},
doi = {10.1364/OME.394724},
}

\end{document}